\documentclass[%
 reprint,
superscriptaddress,
 amsmath,amssymb,
aps,
]{revtex4-2}

\usepackage{graphicx}% Include figure files
\usepackage{dcolumn}% Align table columns on decimal point
\usepackage{bm}% bold math

\usepackage{amsmath}

\usepackage{color}

\begin{document}

\preprint{APS/123-QED}

\title{Unmasking charge transfer in the Misfits: ARPES and \emph{ab initio} prediction of electronic structure in layered incommensurate systems without artificial strain
}

\author{Drake Niedzielski}
\affiliation{%
 Department of Physics, Cornell University.
}
\author{Brendan D.~Faeth}
\affiliation{%
 Platform for the Accelerated Realization, Analysis, and Discovery of Interface Materials (PARADIM), Cornell University.
}
\author{Berit H.~Goodge}
\affiliation{%
 School of Applied and Engineering Physics, Cornell University.
}
\affiliation{%
*current address: Max Planck Institute for Chemical Physics of Solids, 01187 Dresden, Germany.
}

\author{Mekhola Sinha}
\affiliation{%
 Department of Chemistry, The Johns Hopkins University.
}
\author{Tyrel M.~McQueen}
\affiliation{%
 Department of Chemistry, The Johns Hopkins University.
}
\affiliation{%
Institute for Quantum Matter, William H.~Miller III Department of Physics and Astronomy, The Johns Hopkins University.
}
\affiliation{%
Department of Materials Science and Engineering, The Johns Hopkins University.
}

\author{Lena F.~Kourkoutis}
\affiliation{%
 School of Applied and Engineering Physics, Cornell University.
}
\affiliation{%
 Kavli Institute at Cornell for Nanoscale Science, Cornell University, Ithaca, NY 14850
}

\author{Tomás A.~Arias}
\affiliation{%
 Department of Physics, Cornell University.
}

\date{\today}% It is always \today, today,
             %  but any date may be explicitly specified

\begin{abstract}
Common belief is that the large band shifts observed in incommensurate misfit compounds, e.g. (LaSe)$_{1.14}$(NbSe$_2$)$_{2}$, are due to interlayer charge transfer. In contrast, our analysis, based on both ARPES measurements and a specialized \emph{ab initio} framework employing only quantities well defined in incommensurate materials, demonstrates that the large band shifts instead reflect changes in valence band hybridization and interlayer bonding.
The strong alignment of our \emph{ab initio} predictions and ARPES measurements confirms our understanding of the incommensurate electronic structure and charge transfer.
\end{abstract}

\maketitle

The unexpected discovery of superconductivity in twisted bilayer graphene \cite{cao_unconventional_2018} has lead to the new field of incommensurate Moir\'e materials \cite{carr_electronic-structure_2020, andrei_marvels_2021, mak_semiconductor_2022}. Bulk incommensurate materials have also been synthesized, with the so-called ``misfits'', having been first grown in the 1970s and characterized in the 1980s \cite{takahashi_synthesis_1971, makovicky_non-commensurate_1981, wiegers_structure_1988}. The structure of misfits is particularly intriguing, comprising one or more layers of a transition metal dichalcogenide (TMD) stacked with alternating layers of a rare earth rock salt. The TMDs themselves host many interesting electronic phases \cite{manzeli_2d_2017,zhou_ising_2016}, which can then be heavily doped by the adjacent rock salt layers \cite{yao_charge_2018,leriche_misfit_2021,zhong_revealing_2023}, making the misfits a promising platform for manipulating exotic electronic phases \cite{ng_misfit_2022}.

Despite recent interest in this doping approach, the existence, extent, and nature of charge transfer effects in various misfit compounds have been subjects of contention, as summarized by G\"{o}hler \emph{et al.}~\cite{gohler_charge_2018}. Early studies concluded that no charge transfer could be present in a wide range of misfit compounds, due to negligible shifts in core-level electron energies \cite{ettema_leed_1992,a_r_h_f_ettema_x-ray_1993}. Later studies suggested that some charge transfer must occur \cite{ohno_interlayer_1991,gohler_charge_2018}, but rarely provided quantitative estimates of the transferred charge. Recent angle-resolved photoemission spectroscopy (ARPES) studies, on the other hand, have observed significant band shifts near the Fermi level, indicating substantial doping \cite{yao_charge_2018,leriche_misfit_2021,zhong_revealing_2023,chikina_one-dimensional_2022}, which is widely interpreted as an interlayer charge transfer into the TMD \cite{fang_electronic_1995, g_a_wiegers_electrical_1990,samuely_extreme_2021,d_berner_lase114nbse22_1997,zullo_misfit_2023,leriche_misfit_2021,chikina_one-dimensional_2022,zhong_revealing_2023}. Finally, 
recent progress in understanding this extensive doping has revealed correlation between the large band shifts and the work functions of the two materials \cite{zullo_misfit_2023}, but falls short of quantitative understanding of the expected transfer and links band doping directly to charge transfer without justification.

In terms of theory, tight-binding methods can predict detailed electronic structures and can be applied to extremely large approximate model systems \cite{fang_ab-initio_2015, massatt_electronic_2017, carr_exact_2019}. However, tight-binding is not fully \emph{ab initio}, requires significant upfront effort to model new materials, and, unless specially formulated, is not adept at accounting for charge transfer. In contrast, performing fully \emph{ab initio} density-functional theory (DFT) calculations using approximate supercells with large inbuilt strains result in an inherent trade-off between accuracy and computational resources \cite{leriche_misfit_2021}.

To address the above challenges, this Letter proposes a general formalism for the \emph{ab initio} treatment of incommensurate materials by working with microscopic quantities that remain well-defined even in the absence of translational symmetry. Moreover, to compute these quantities \emph{ab initio}, we generalize the Mismatched INterface Theory (MINT) of Gerber \emph{et al.} \cite{gerber_ab_2020} to periodically stacked systems and describe a general method for computing the electronic structure and phonon dispersion of incommensurate materials. We then deploy the approach to predict the \emph{ab initio} 
charge transfer, doping, and ARPES spectra of a material \emph{with no underlying periodicity}, specifically the (LaSe)$_{1.14}$(NbSe$_2$)$_{2}$ misfit compound. We then validate our approach through our own ARPES measurements, finding close agreement between the measured and predicted doping of the Fermi-level crossing bands, indicating that we have an accurate \emph{ab initio} understanding of the detailed electronic structure of the material. Our \emph{ab initio} results further predict an order-of-magnitude discrepancy between the doping in the Fermi-level crossing bands and the actual interlayer charge transfer, exposing a gap in the previous understanding of misfit materials. Finally, we resolve this apparent discrepancy through careful analysis of orbital-projected band structures and changes in the real-space electron density.

\emph{Formalism ---} Almost all electronic quantities within \emph{ab initio} density-functional theory calculations are extractable from the one-electron spectral weight function \cite{hedin_beyond_1970, hedin_effects_1970}, which remains well defined in the absence of any underlying periodicity. In its Fourier representation (of particular interest for ARPES), this function is
\begin{equation}
 \rho(q,q',\epsilon) = \sum_i \tilde{\psi_{i}}(q) \tilde{\psi}_{i}^*(q') \delta(\epsilon-\epsilon_i), 
 \label{eqn:rho_e}
\end{equation}
where, following common practice, we approximate the quasi-particle energies and orbitals with the Kohn-Sham eigenvalues and orbitals  $\epsilon_i,\psi_i$ \cite{kohn_self-consistent_1965}. When there is translational symmetry, the index $i$ decomposes into the standard band/crystal-momentum indices $nk$, with the $\psi_{nk}$ becoming Bloch waves. Likewise, nearly all vibrational properties are expressible from
\begin{equation}
\pi(q,q',\omega) = \sum_a \tilde{\vec{\phi}}_a(q) \tilde{\vec{\phi}}_a^\dag(q') \delta(\omega-\omega_a),
\label{eqn:rho_ph}
\end{equation}
where $\vec{\phi}_{a} (r) = \sum_\tau \vec{\xi}^\tau_{a} \delta(r-r_\tau)$ is the phonon displacement field and $\tilde{\vec{\phi}}_a(q)$ its Fourier transform. Analogously, when there is translational symmetry, the state index $a$ factors into the branch-index/crystal-momentum indices $\alpha k$. Even the intricate Eliashberg spectral function, used in \emph{ab initio} calculations of superconducting transition temperatures, is expressible in terms of Eqs.~(\ref{eqn:rho_e},\ref{eqn:rho_ph}),
\begin{equation} \label{eqn:eliashberg}
\begin{aligned}
\alpha^2 F(\omega,\mu) = \frac{1}{N(\mu)} \int dq_1\, dq_2\, dq_3\, dq_4 \ \rho(q_4,q_1,\mu) \rho(q_2,q_3,\mu) \\
\Delta \tilde{\vec{V}}^\top(q_1-q_2) \pi(q_1-q_2,q_3-q_4, \omega)  \Delta \tilde{\vec{V}}^*(q_3-q_4),
\end{aligned}
\end{equation}
where $\tilde{\vec{V}}$ is the Fourier transform of the atom-displacement perturbation potential, and $N(\mu)\equiv\int dq\,\rho(q,q,\mu)$ is the Fermi-level density of states.

To obtain the spectral functions Eqs.~(\ref{eqn:rho_e},\ref{eqn:rho_ph}) for incommensurate systems, we propose \emph{MINT-Sandwich} to extend the method of Gerber \emph{et. al.} \cite{gerber_ab_2020}, which was limited to 2D material interfaces. We now consider fully incommensurate 3D systems as the limit of a sequence of prototype periodic materials with unit cells of increasing size that contain small (approximately) constant gaps in one of the two materials to accommodate the incommensurability. Due to ``nearsightedness''  \cite{kohn_density_1996, ismail-beigi_locality_1999, prodan_nearsightedness_2005}, physical observables, computed as integrals of the products of the spectral functions Eqs.~(\ref{eqn:rho_e},\ref{eqn:rho_ph}) with smooth functions such as matrix elements (\emph{e.g.},  Eq.~(\ref{eqn:eliashberg})), will converge over the prototype sequence to the true values for the incommensurate system. The observable properties of the incommensurate material can be extracted from their asymptotic behavior over the prototype sequence, without the need to navigate between the Scylla of artificial strain and the Charybdis of extremely large supercells at accidental approximate lattice matches.

\begin{figure}
\includegraphics[width=\linewidth]{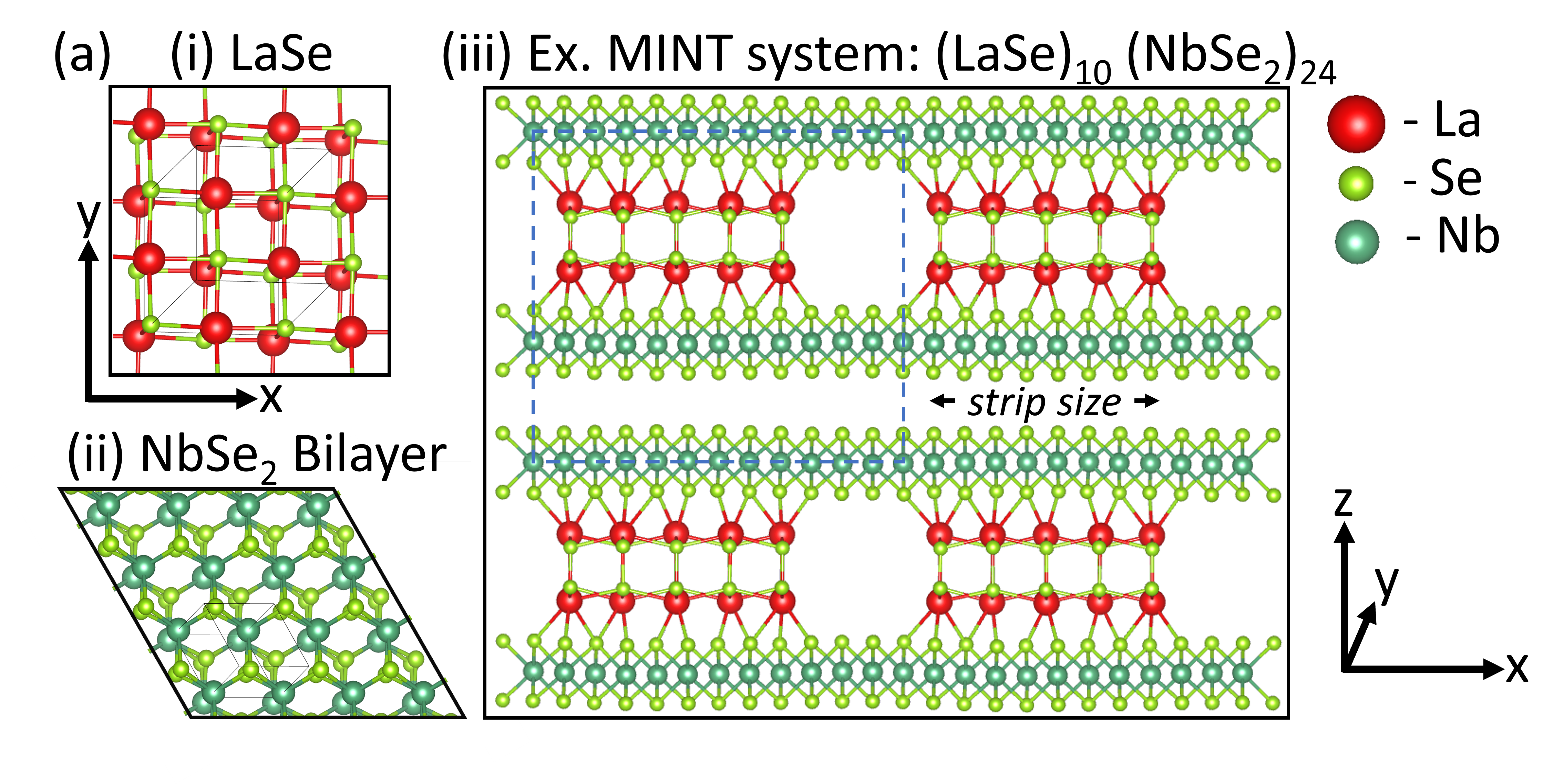}
\includegraphics[width=\linewidth] {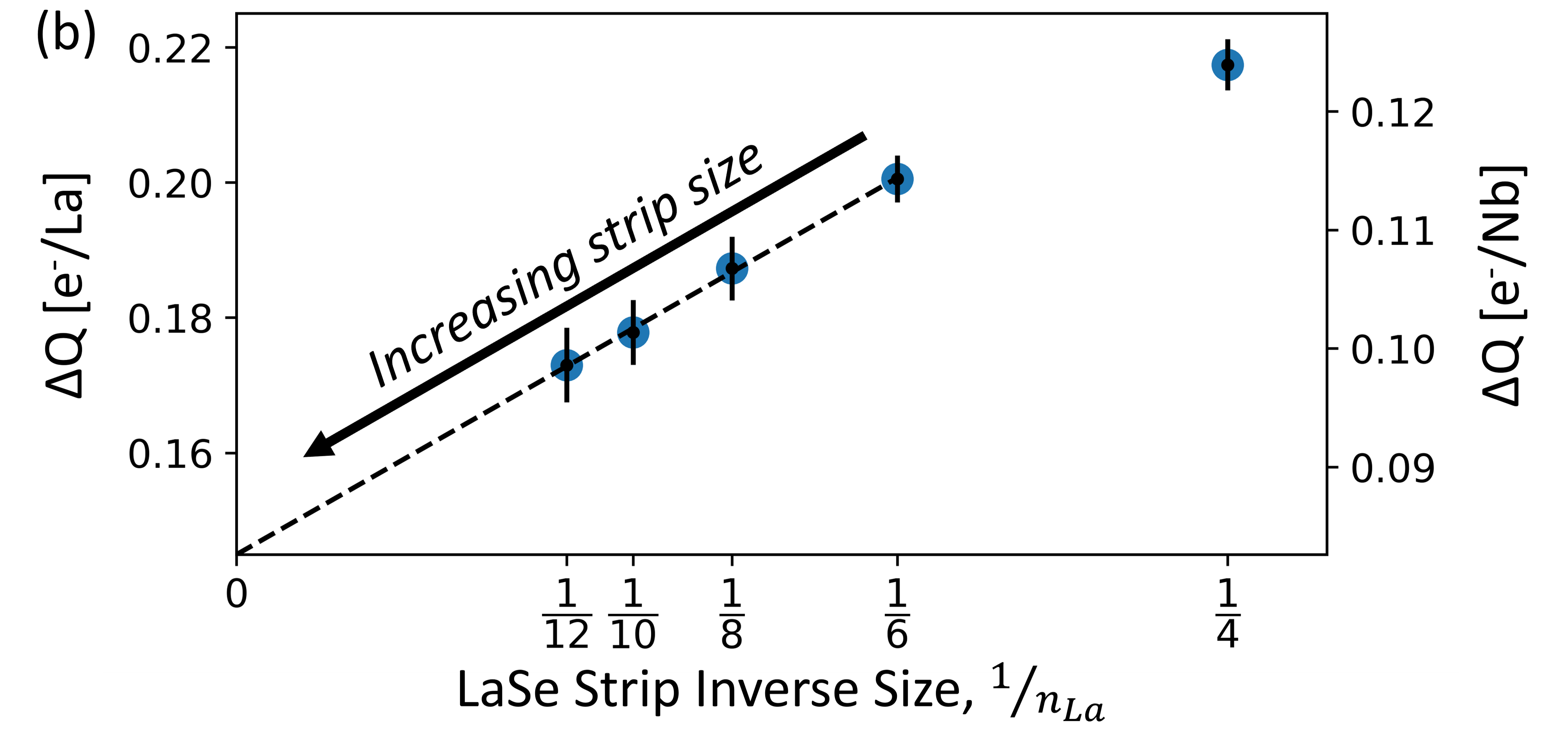}
\caption{(a) Top view of constituent, incommensurate LaSe rocksalt (i) and 2H-NbSe$_2$ bilayer (ii), and side view of $n_\mathrm{La} = 10$ MINT-Sandwich periodic prototype (iii) with supercell indicated by blue-dashed line. The LaSe square lattice (i) and the NbSe$_2$ hexagonal lattice (ii) lead to a 1:1 commensurate lock-in in the \emph{y}-direction, but an incommensurate mismatch $\sim$1.14:1 in the \emph{x}-direction. Gaps between periodic images of the LaSe strip prevent artificial strain. (b) Interlayer charge transfer $\Delta Q$ versus inverse strip size for the MINT-Sandwich prototype sequence, with best-fit linear relationship (dashed line).
}
\label{fig:MatEx}
\end{figure}

\emph{Sequence construction ---} Figure \ref{fig:MatEx}(a)(iii) shows one member of a sequence which converges to (LaSe)$_{1.14}$(NbSe$_2$)$_2$. This sequence represents each NbSe$_2$ substrate with an $(n_\mathrm{Nb}/2) \times \sqrt{3}$ rectangular supercell containing $n_\mathrm{Nb}$ niobium atoms. We found the sequence $n_\mathrm{Nb}=6,8,10,12,14,\ldots$ to converge at $n_\mathrm{Nb}=14$ for our quantities of interest. We further exploited the known lattice match between the $[010]_{\text{NbSe}_2}$ and $[\overline{1}10]_{\text{LaSe}}$ directions ($y$-direction, Figure~\ref{fig:MatEx}(a)) to treat each LaSe cluster as periodic in that direction, while including gaps of approximately constant size ($7.6-9.4$ \AA) in the orthogonal direction to accommodate the incommensurability. The resulting unit cells contain $n_\mathrm{La}=4,6,8,10,12$ lanthanum atoms, respectively. Finally, we account for the 3D structure by layering periodically in the out-of-plane direction ($z$, Figure~\ref{fig:MatEx}(a)). (Further computational details are in SM and references \cite{sundararaman_jdftx_2017,garrity_pseudopotentials_2014,perdew_generalized_1996}.)

\emph{Charge Transfer ---} For each member of our MINT-Sandwich sequence, we compute the total charge transfer from the electron density, $n_\text{misfit}(r) \equiv \int  \rho(r,r,\epsilon) f(\epsilon)\, d\epsilon$, where $f(\epsilon)$ is the Fermi function. Subtracting the densities of the isolated LaSe and NbSe$_2$ layers from the combined system and integrating over the NbSe$_2$ layer yields the charge transfer, which we then average over multiple different relative shifts of the two layers. (See SI for details.)
% $0.08265 \pm 0.0006$e$^-$/Nb
Plotting interlayer charge transfer $\Delta Q$ versus inverse MINT-strip size (Figure~\ref{fig:MatEx}(b)) produces a linear relation \cite{gerber_ab_2020} whose vertical intercept gives the limit of our prototype sequence, yielding a final \emph{ab initio} interlayer charge transfer for (LaSe)$_{1.14}$(NbSe$_2$)$_2$ of $0.145 \pm 0.001$~e$^-$/La = $0.0827 \pm 0.0006~$e$^-$/Nb from the LaSe to NbSe$_2$ layers. 

As a cross-check, we compare our result to Luryi's quantum capacitor model \cite{luryi_quantum_1988}, 
\begin{equation}
 \Delta Q = \bigg[ \frac{1}{C_{\text{Geo}}} + \frac{1}{g_{\text{LaSe}}e^2} + \frac{1}{g_{\text{NbSe}_2}e^2} \bigg]^{-1} \frac{ \mu_{\text{LaSe}} - \mu_{\text{NbSe}_2} }{e} , 
\end{equation}
where we use the \emph{ab initio} Fermi-levels of the isolated infinite layers ($\mu_{\text{LaSe}} = -2.75$~eV, $\mu_{\text{NbSe}_2}= -5.54$~eV), their \emph{ab initio} quantum capacitances ($g_{\text{LaSe}} = 0.102~$eV$^{-1}$\AA$^{-2}$, and $g_{\text{NbSe}_2} = 0.115~$eV$^{-1}$\AA$^{-2}$), and the geometric capacitance derived from the interlayer distance ($C_{\text{geo}} = 4.54\times10^{-3}~$e$^2\cdot\,$eV$^{-1}$\AA$^{-2}$). The result is 0.111~e$^-$/La = 0.0635~e$^-$/Nb, in good agreement with our \emph{ab initio} result considering the semi-empirical nature of Luryi's model. This agreement indicates that our approach captures the underlying charge transfer physics, but now from a fully \emph{ab initio} perspective.

\begin{figure}
\includegraphics[width=\linewidth] {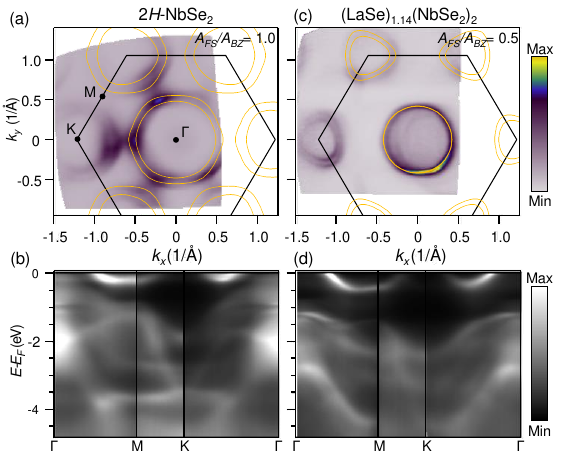}
\caption{ARPES measured Fermi surfaces (top row) and valence band structure (bottom row) for undoped 2H-NbSe$_2$ (a,b) and (LaSe)$_{1.14}$(NbSe$_2$)$_2$ ``misfit'' crystals (c,d).  (a,c): Fermi surfaces are generated from a series of constant energy contours integrated over $\pm$5~meV about the Fermi level and combined to generate the complete mapping, with black lines indicating the NbSe$_2$ hexagonal Brillioun Zone boundary and orange lines depicting Fermi-surface contours derived from best fits to a tight binding model for both data sets.}
\label{fig:ARPES}
\end{figure}

%%%%%%%%%%%%%%%%%%%%%%%%%%%%%%
%Begin ARPES 
Next, we compare our predicted interlayer charge transfer against the results of our ARPES experiments (see SI for details). Figure~\ref{fig:ARPES} presents ARPES measurements for both undoped 2H-NbSe$_2$ (a,b) and (LaSe)$_{1.14}$(NbSe$_2$)$_2$ ``misfit" crystals (c,d). 
For bulk 2H-NbSe$_2$, the Fermi surface (Figure~\ref{fig:ARPES}(a)) and valence band structure (Figure~\ref{fig:ARPES}(b)) are consistent with previous reports \cite{rahn_gaps_2012,bawden_spinvalley_2016}. The Fermi-level crossing bands form a set of hole-like pockets with predominantly Nb 4d character centered at both $\Gamma$ and $K$, along with a much fainter and more diffuse k$_z$-dependent Se 4p band closer to $\Gamma$. The Fermi surface maps for both of the Nb-derived pockets show clear splitting arising from interlayer interactions in the 2H structure \cite{bawden_spinvalley_2016}.  For (LaSe)$_{1.14}$(NbSe$_2$)$_2$, the Fermi surface more resembles that of monolayer NbSe$_2$ \cite{nakata_anisotropic_2018}, where the $\Gamma$-centered pocket is nearly degenerate but with inversion-symmetry breaking leading to a slight spin-splitting in the K-centered pocket \cite{nakata_anisotropic_2018}.  These splittings are less evident in the high-symmetry scans (Figure~\ref{fig:ARPES}(b,d)) than the Fermi surfaces (Figure~\ref{fig:ARPES}(a,c)) due to differences in measurement geometries and resulting photoemission matrix elements.

The clear relative shift in the (LaSe)$_{1.14}$(NbSe$_2$)$_2$ bands to higher binding energies indicates a large effective electron doping in the NbSe$_2$ layer of the misfit compound. To quantify this, we perform a Luttinger volume analysis with the dispersion of the Fermi-level crossing bands described with a tight-binding model. We fit this model against the ARPES data following a scheme described in more detail elsewhere \cite{rahn_gaps_2012,rossnagel_fermi_2005} to generate Fermi surface contours for both samples (orange lines, Figures~\ref{fig:ARPES}(a,c)). The change in area of the resulting hole-like pockets yields a doping of $0.50 \pm 0.035$~e$^-$/Nb in (LaSe)$_{1.14}$(NbSe$_2$)$_2$, in good agreement with the 0.55-0.6~e$^-$/Nb found by Leriche \emph{et al.} using a different analysis technique \cite{leriche_misfit_2021}. 
At first glance, these ARPES measurements \emph{appear} to be in strong disagreement with both the quantum capacitor model and our \emph{ab initio} result of $0.083$~e$^-$/Nb, reminiscent of the aforementioned disagreement on misfit charge transfer in the literature \cite{gohler_charge_2018,ettema_leed_1992, a_r_h_f_ettema_x-ray_1993, yao_charge_2018,leriche_misfit_2021,zhong_revealing_2023}.
% End ARPES
%%%%%%%%%%%%%%%%%%%%%%%%%%%%%%%%%%%%%%%%%%%%%%%%%% 

\begin{figure*}
\includegraphics[width=0.99\linewidth]{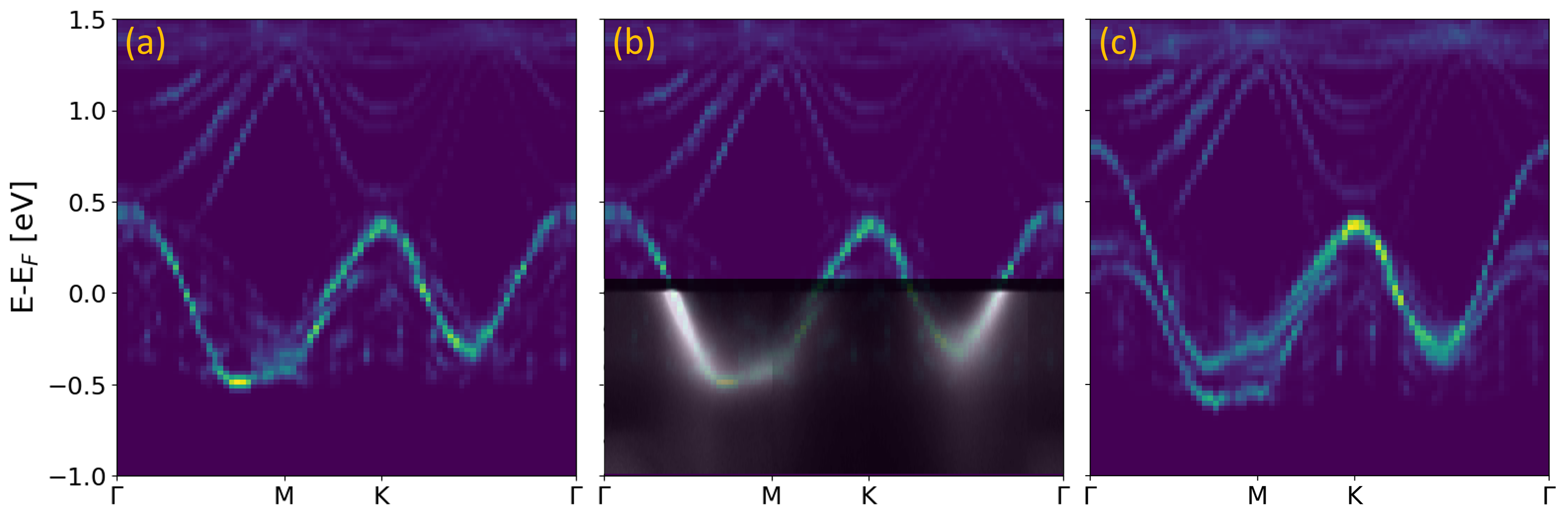}
\caption{ \emph{Ab initio} spectral intensity plots versus distance $\epsilon=E-E_f$ from Fermi-level (vertical) and wave vector $q$ (horizontal): brighter hues signify greater spectral intensity $n(q,\epsilon)$. 
(a) MINT-Sandwich prediction for (LaSe)$_{1.14}$(NbSe$_2$)$_2$ surface showing a single distinct band crossing the Fermi level, in accordance with experimental ARPES data. (b) Overlay of experimental ARPES data with result from (a) showing exceptional agreement between theory and experiment. (c) Prediction for bulk (LaSe)$_{1.14}$(NbSe$_2$)$_2$ showing two Fermi-level crossing bands reminiscent of bulk NbSe$_2$.} 

\label{fig:ARPES:DFT}
\end{figure*}

\emph{MINT-Sandwich calculation of ARPES signature ---} We next predict the expected ARPES signature directly from exactly the \emph{same} \emph{ab initio} calculations that led to the apparently discrepant charge-transfer prediction. Lack of lattice-translation symmetry in incommensurate materials poses a fundamental challenge to ARPES prediction: without a Brillouin zone, the band structure is no longer well-defined. Fortunately, apart from matrix-element corrections, ARPES actually probes the momentum-resolved density of states (DOS), $n(q,\epsilon) \equiv \rho(q,q,\epsilon)$, which remains well-defined even in the absence of translational symmetry.

To calculate $n(q,\epsilon)$ for an incommensurate system, we begin by using Wannier function techniques to compute $n^{(M)}(q,\epsilon) = \sum_n \int_{SBZ} dk' |\psi_{nk'}(q)|^2 \delta(\epsilon-\epsilon_{nk'})$ for each periodic member $M$ of the MINT-Sandwich sequence, where $SBZ$ is the corresponding supercell Brillouin zone. Then, to allow direct comparison with our experimental results, we fold the resulting $n^{(M)}(q,\epsilon)$ into the NbSe$_2$ BZ, $n^{(M)}_{BZ(\textrm{NbSe}_2)} (k,\epsilon)  = \sum_g n^{(M)}(k+g,\epsilon)$, using techniques adapted from \cite{mayo_band_2020}. Next, to accelerate convergence, we weight the states by their corresponding probabilities within the regions that include the LaSe strips (details in Supplemental Information). The resulting predictions show significant changes when going from LaSe strips of width $n_\mathrm{La} = 4$ to $n_\mathrm{La} = 6$, with minor changes as $n_\mathrm{La}$ increases further. The final change between $n_\mathrm{La} = 10$ and $n_\mathrm{La} = 12$  is visually imperceptible. All ARPES results below are thus are reported for $n_\mathrm{La}=12$.

\emph{Results ---} Figure~\ref{fig:ARPES:DFT} displays our strain-free \emph{ab initio} MINT-Sandwich ARPES predictions for (LaSe)$_{1.14}$(NbSe$_2$)$_2$.
%TAA: HERE at 2024-04-24-11:57
%Planning to join these paragraphs
Because ARPES is surface sensitive and the NbSe$_2$ interlayer Van der Waals bonds break most easily during exfoliation \cite{leriche_misfit_2021}, Figure~\ref{fig:ARPES:DFT}(a) first presents our results for an (LaSe)$_{1.14}$(NbSe$_2$)$_2$ surface terminated with a single NbSe$_2$ monolayer. Figure~\ref{fig:ARPES:DFT}(a) predicts that a single discernible band crosses the Fermi level, consistent with our experimental data (Figure~\ref{fig:ARPES}(d)), apart perhaps from the small spin-splitting evident in Figure~\ref{fig:ARPES}(c) and not considered in our \emph{ab initio} calculations. Figure~\ref{fig:ARPES:DFT}(b) overlays Figures~\ref{fig:ARPES:DFT}(a) on \ref{fig:ARPES}(d), revealing a remarkable agreement for both the shape and energies of our calculated and experimental Fermi-level crossing bands, affirming the reliability of our MINT-Sandwich computational framework. Finally, Figure \ref{fig:ARPES:DFT}(c) shows our prediction for the electronic structure of incommensurate bulk (LaSe)$_{1.14}$(NbSe$_2$)$_2$, which exhibits two clearly discernible bands crossing the Fermi level, reminiscent of 2H-NbSe$_2$ (Figure \ref{fig:ARPES}(a)), but in sharp contrast to the single band observed for the (LaSe)$_{1.14}$(NbSe$_2$)$_2$ surface.

\begin{figure*}
\begin{tabular}{ c c l }
\hphantom{MM} (a) Isolated (LaSe)$_6$ \hphantom{MMMM} (b) Isolated (NbSe$_2$)$_{16}$  \hphantom{MMM} (c) (LaSe)$_6$(NbSe$_2$)$_{16}$ & (d) Valence & \hphantom{M}(e) Fermi \\
\includegraphics[width=0.75\linewidth]{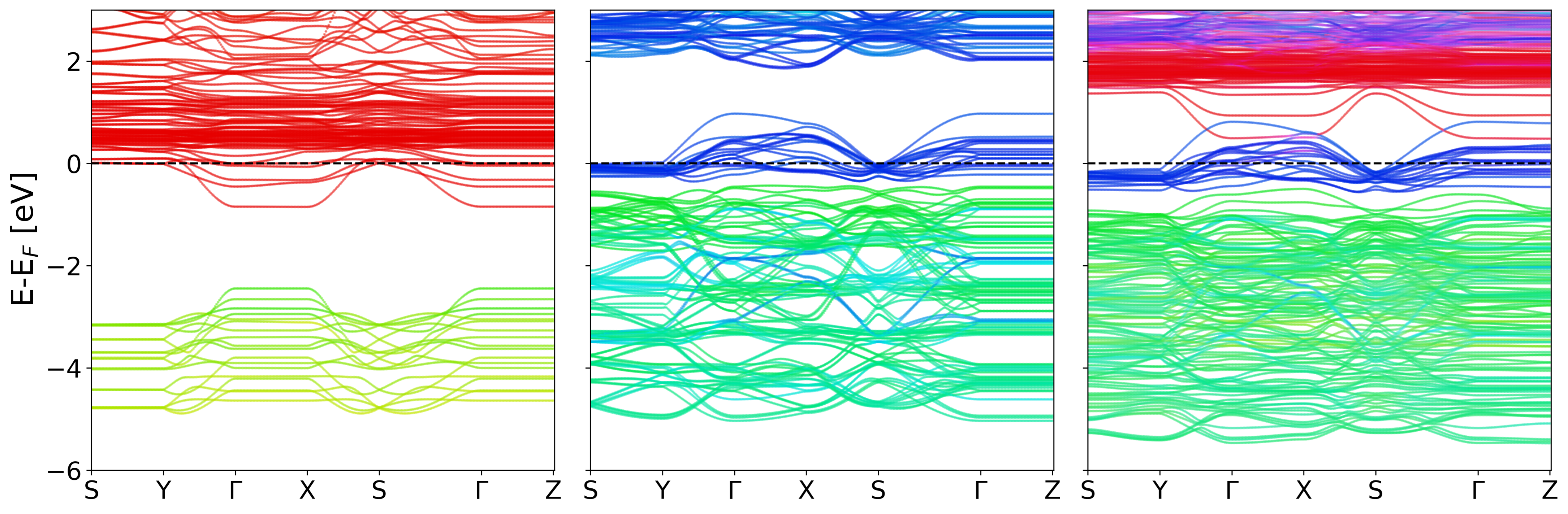} &
\includegraphics[width=0.125\linewidth]{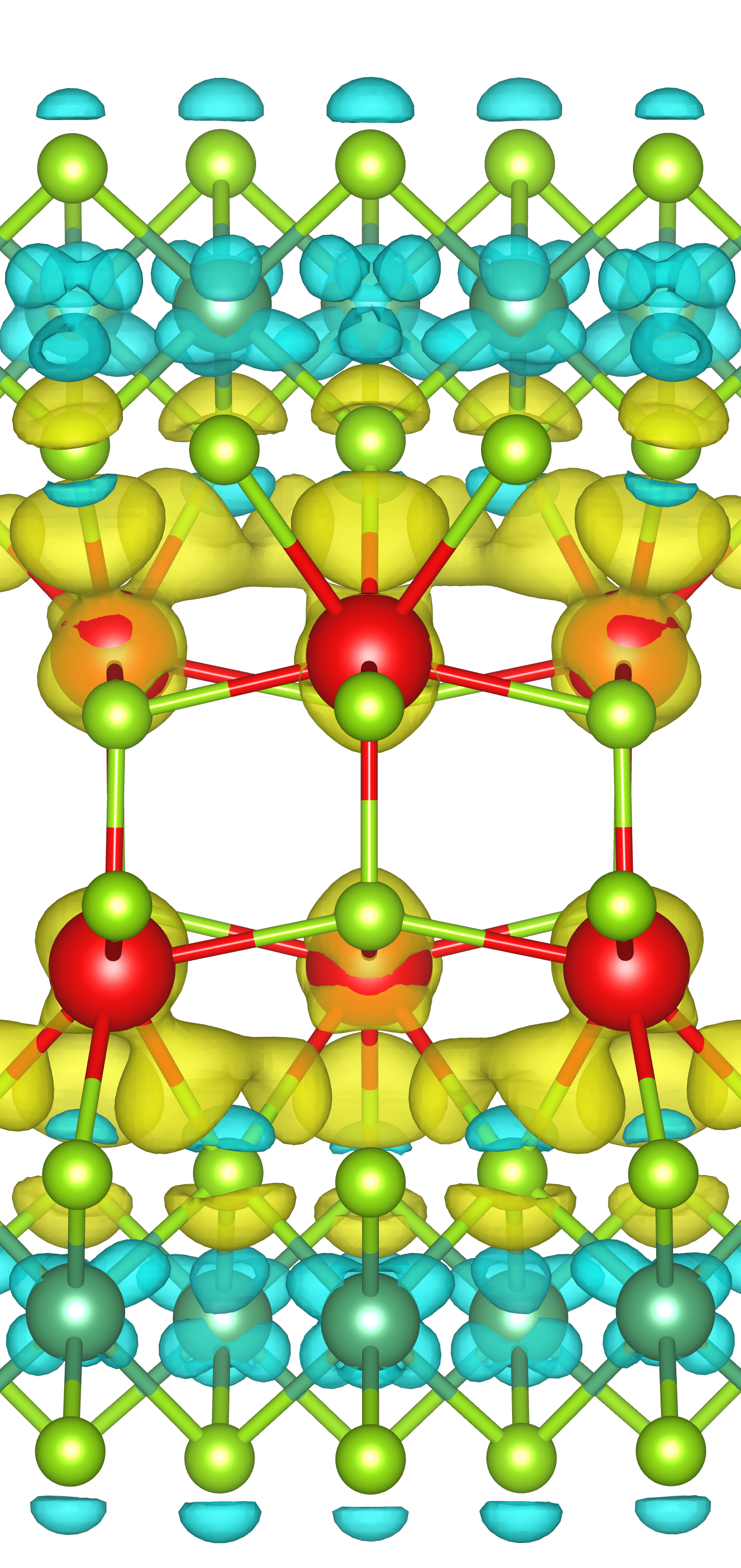} & \includegraphics[width=0.125\linewidth]{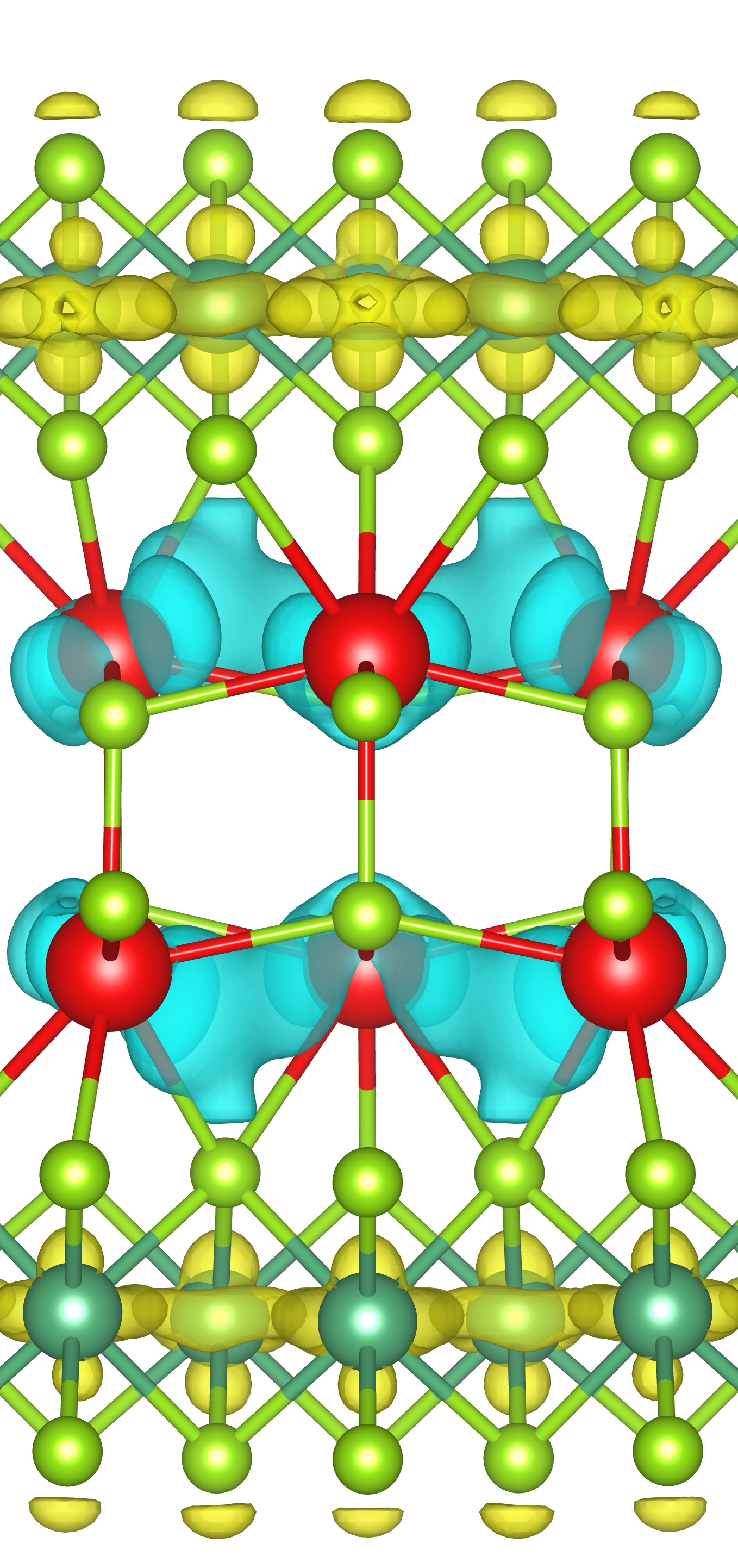} 
\end{tabular}
\caption{Charge transfer and doping in (LaSe)$_{1.14}$(NbSe$_2$)$_2$ in momentum space (a-c) and real space (d,e) for $n_\mathrm{La}=6$ member of MINT-Sandwich prototype sequence. (a-c): band structures of (a) the isolated LaSe strip, (b) the NbSe$_2$ substrate, and (c) their bonded combination, with each band rgb color coded in proportion to its projections onto the different atomic species (La=red,Se=green,Nb=blue, as in Figure~\ref{fig:MatEx}). Transparency is used so that overlapping bands do not obscure each other.  (d,e): Changes in real-space electron density during the bonding process for (d) the completely filled valence bands and (e) the Fermi-level crossing bands, with yellow/blue contours indicating regions of increasing/decreasing density and the atoms following the same conventions as Figure~\ref{fig:MatEx}.
}
\label{fig:bandProj}
\end{figure*}

\emph{Resolution of \emph{apparent} charge discrepancy ---} 
Our ability to predict the same large shift (Figure~\ref{fig:ARPES:DFT}(b)) of the Fermi-level crossing bands as observed in ARPES indicates that our calculations indeed predict a similarly large doping into those bands ($\sim$0.5~e$^-$/Nb). The small interlayer charge transfer ($\sim$0.08~e$^-$/Nb) in these very same calculations leads to the conclusion that \emph{the charge doping into the d-bands must not be the same as the net interlayer charge transfer}. Indeed, Shao \emph{et al.} noted a similar discrepancy between the shifts in Fermi-level crossing bands and the net interlayer charge transfers predicted from a capacitor model for TaS$_2$ on a Au(111) substrate \cite{shao_pseudodoping_2019}. Those authors then proposed a new mechanism, \emph{pseudodoping}, where increased hybridization of bands near the Fermi surface allows greater band movement while keeping net charge transfer small. We find that (LaSe)$_{1.14}$(NbSe$_2$)$_2$ exhibits a related, more subtle phenomenon driven by hybridization changes \emph{far} from the Fermi surface.

Figure~\ref{fig:bandProj}(a-c) shows, for our $n_\mathrm{La} = 6$ MINT proxy system, the predicted band hybridization change going from the two isolated layers to their interacting misfit combination.
Going from the isolated LaSe layer (Figure~\ref{fig:bandProj}(a)) to the combined (LaSe)$_{1.14}$(NbSe$_2$)$_2$ system (Figure~\ref{fig:bandProj}(c)), the La bands (red), some of which are filled in the isolated layer, completely empty out. These bands thereby donate $\sim$1~e$^-$/La to the Nb bands (blue) that cross the Fermi level, as evident from the significant downward shift of these Fermi-level crossing Nb-\emph{d} bands  (blue)  when going from the isolated NbSe$_2$ system (Figure~\ref{fig:bandProj}(b)) to the combined system (Figure~\ref{fig:bandProj}(c)).  This transfer
explains the substantial charge doping of $\sim$0.5~e$^-$/Nb observed in ARPES. However, the observed decrease in Nb character (blue, cyan) of the lower valence bands at $\sim$2~eV below the Fermi level, when going from the isolated NbSe$_2$ layer (Figure~\ref{fig:bandProj}(b)) to the combined system (Figure~\ref{fig:bandProj}(c)), largely offsets this charge movement, explaining the small \emph{net} interlayer charge transfer that we predict.

Mechanistically, we propose that the change in valence-band hybridization is what drives the large doping into the Nb-\emph{d} bands, in what amounts to a \emph{polarized} quantum capacitor model. Figure~\ref{fig:bandProj}(d) shows the movement of the bound charge associated with the fully filled valence bands (mostly green/yellow with some blue shading). As the two layers interact, this charge accumulates between the lanthanum atoms of the LaSe layer and the selenium atoms of the NbSe$_2$ layer, indicating formation of interlayer covalent bonds, inducing a large bound-charge polarization between the La atoms and the Nb layer.
Next, because these La and Nb layers have bands at the Fermi level, these bands then shift rigidly to screen this polarization by adjusting their charges. (Figure~\ref{fig:bandProj}(e) shows the shift in electron density associated with these bands.) With perfect screening, interlayer charge transfer would be zero; however, the original work-function difference between the layers results in the final small net interlayer charge transfer.

Finally, we directly compute the \emph{ab initio} charge doping into the Fermi-level crossing Nb-\emph{d} bands within MINT-Sandwich. Using only the electron densities from Fermi-level crossing bands (Figure \ref{fig:bandProj}(e)), we find, after finite size extrapolation, an \emph{ab initio} value of 0.921~e$^-$/La = 0.525~e$^-$/Nb for incommensurate (LaSe)$_{1.14}$(NbSe$_2$)$_2$. This is in even closer agreement with our predicted ARPES value of $0.50 \pm 0.035$~$e^-$/Nb than a previous \emph{ab initio} study \cite{leriche_misfit_2021} that employed  a commensurate supercell ($\sim$0.55-0.6~$e^-$/Nb).

\emph{Conclusion ---} MINT-Sandwich, the extension of MINT to 3D systems, enables the \emph{ab initio} study of important quantities including net charge transfers, ARPES, and other spectral functions in \emph{aperiodic} systems without the introduction of artificial strain. With MINT-Sandwich, we have resolved the apparent discrepancy between the net interlayer charge transfer and the ARPES measured d-band doping in (LaSe)$_{1.14}$(NbSe$_2$)$_2$. We show that the doping of the TMD d-bands is due not to a net interlayer charge transfer, as previously believed, but to polarization induced from restructuring of interfacial covalent bonds. This new understanding opens the path to the rational design of incommensurate material interfaces and heterostructures with heavy effective doping in the misfits and beyond.

\emph{Acknowledgements ---} This work made use of the theory, thin film growth, electron microscopy, and bulk crystal growth facilities of the Platform for the Accelerated Realization, Analysis, and Discovery of Interface Materials (PARADIM), which are supported by the National Science Foundation under Cooperative Agreement No. DMR-2039380. D.~Niedzielski was supported full time and T.A.~Arias acknowledges partial support under the same agreement.

\bibliographystyle{apsrev4-1}
\bibliography{bibliography}

\end{document}